\documentclass[aps,prl,twocolumn,superscriptaddress,groupedaddress]{revtex4}
\usepackage{graphicx}  
\usepackage{dcolumn}   
\usepackage{bm}        
\usepackage{amssymb}   

\begin{document}

\title{A Hermetic On-Cryostat Helium Source for Low Temperature Experiments}
\author{K.~E.~Castoria}
\affiliation{EeroQ Corporation, Chicago, Illinois, USA}
\author{H.~Byeon}
\affiliation{EeroQ Corporation, Chicago, Illinois, USA}
\author{J.~Theis}
\affiliation{EeroQ Corporation, Chicago, Illinois, USA}
\author{N.~R.~Beysengulov}
\affiliation{EeroQ Corporation, Chicago, Illinois, USA}
\author{E.~O.~Glen}
\affiliation{EeroQ Corporation, Chicago, Illinois, USA}
\author{G.~Koolstra}
\affiliation{EeroQ Corporation, Chicago, Illinois, USA}
\author{M.~Sammon}
\affiliation{EeroQ Corporation, Chicago, Illinois, USA}
\author{S.~A.~Lyon}
\affiliation{EeroQ Corporation, Chicago, Illinois, USA}
\author{J.~Pollanen}
\affiliation{EeroQ Corporation, Chicago, Illinois, USA}
\author{D.~G.~Rees}
\affiliation{EeroQ Corporation, Chicago, Illinois, USA}

\date{\today}

\begin{abstract}
We describe a helium source cell for use in cryogenic experiments that is hermetically sealed $in$ $situ$ on the cold plate of a cryostat. The source cell is filled with helium gas at room temperature and subsequently sealed using a cold weld crimping tool before the cryostat is closed and cooled down. At low temperature the helium condenses and collects in a connected experimental volume, as monitored via the frequency response of a planar superconducting resonator device sensitive to small amounts of liquid helium. This on-cryostat helium source negates the use of a filling tube between the cryogenic volumes and room temperature, thereby preventing unwanted effects such as such as temperature instabilities that arise from the thermomechanical motion of helium within the system. This helium source can be used in experiments investigating the properties of quantum fluids or to better thermalize quantum devices.
\end{abstract}

\maketitle

The liquid and solid phases of helium at low temperatures are paradigms of quantum matter.  Below 2.17~K $^4$He forms a superfluid Bose-Einstein condensate~\cite{Vinen2004} while below 2.5~mK the lighter isotope $^3$He forms a fermionic superfluid with $p$-wave, spin-triplet pairing~\cite{VW1990}. Whilst the bulk superfluids are ideal systems for studying the macroscopic manifestation of quantum mechanical effects, their microscopic properties are often best investigated and controlled with micro- and nano-fabricated devices employed to probe, for example, the influence of nano-scale confinement on the superfluid phases~\cite{heikkinen2021fragility,Varga2022}, or in compact devices for investigating the generation and detection of excitations within the liquid~\cite{woods2023developing}. Also, the high-quality dielectric properties and excellent thermal conductivity of liquid helium make it an appealing material for integration with quantum devices. In fact, liquid helium is already employed to achieve low electron temperatures in solid-state devices~\cite{xia2000ultra,samkharadze2011integrated,jones2020progress,levitin2022cooling}, reduce thermally driven charge noise~\cite{beysengulov2021noise,lucas2023quantum}, and to improve the coherence of superconducting quantum circuits~\cite{lane2020integrating, lucas2023quantum}. Additionally, hybrid devices enabling the control of single electrons trapped above the surface of liquid helium are being actively pursued for applications in quantum information science~\cite{Schuster2010,koolstra2019coupling,PBR2021}. There is therefore strong interest in developing efficient methods by which to integrate liquid helium with condensed matter experiments and quantum devices. 

Traditionally, helium is introduced to a cryogenic experimental cell from an external gas source via a thin capillary filling tube anchored at each thermal stage of the cryostat~\cite{Pobell}. For $^4$He, condensation occurs close to the 4~K stage of the cryostat, and the liquid becomes superfluid where the temperature falls below the critical temperature $T_{c}=2.17$~K. At the experimental volume, which is typically the lowest point in the enclosed system, the helium cools to the cryostat base temperature. While most of the helium resides as ``bulk'' superfluid at this lowest point, a van der Waals film helium coats all surfaces below the critical temperature. The thickness of this film is approximately 30~nm close to the bulk liquid surface, and decreases as $h^{-\frac{1}{3}}$ with increasing height $h$ from the surface~\cite{Pobell}. In addition, liquid helium can become trapped in volumes smaller than the $^4$He capillary length ($l_c \approx 0.5$~mm) as the capillary action counteracts the gravitational force exerted on the liquid.

\begin{figure*}
\includegraphics[scale=0.2]{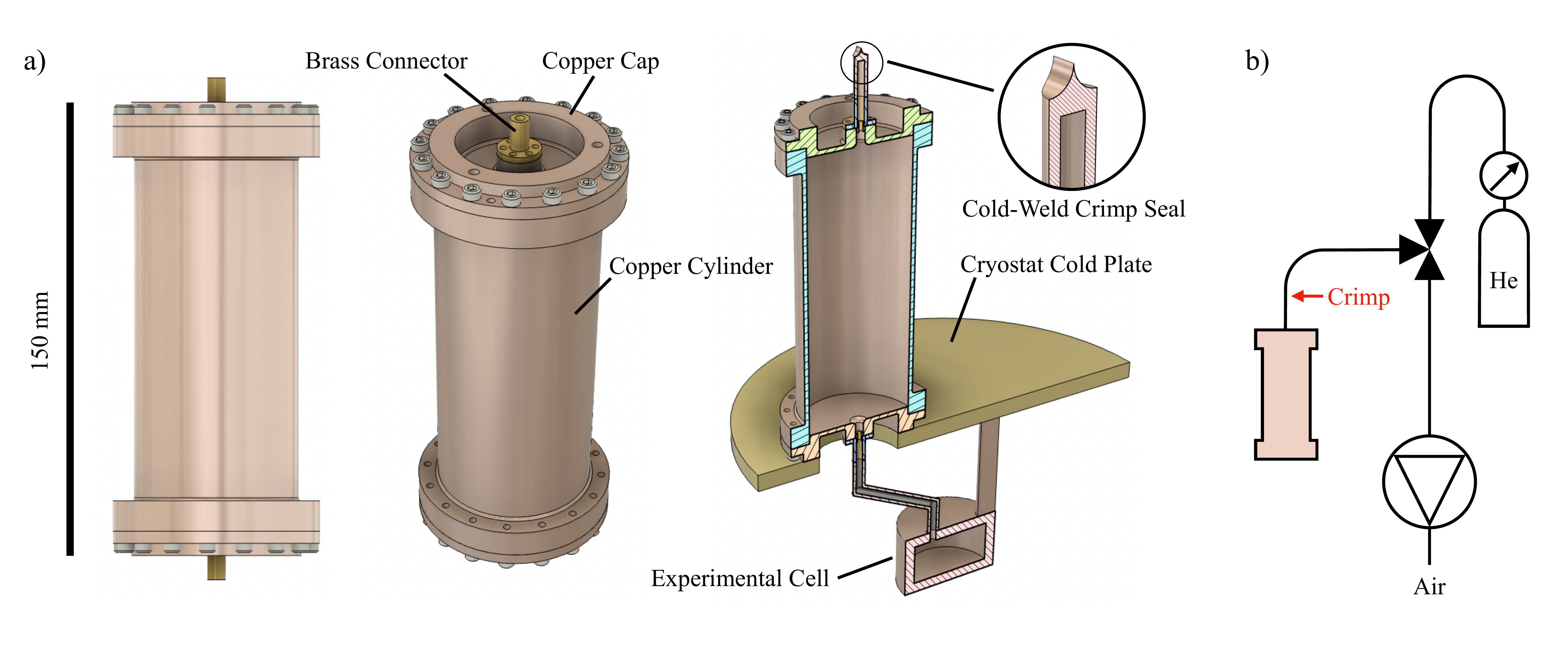}
\caption{\label{fig:epsart} a) Schematic diagram of the hermetic helium source cell, as described in the text, showing (left to right) side view, projected view, and schematic cross section including the experimental cell. b) Schematic of the charging method. A 3-way valve is used to alternate the connection to the source cell between a helium cylinder and a pump.}
\end{figure*}   

The presence of helium in the filling line can give rise to several unwanted effects; uncertainty in the bulk volume is introduced and, as the thermal conductance of superfluid helium is extremely high, heat can be transferred from the higher temperature stages of the cryostat to the base stage. Furthermore, temperature gradients between cryostat stages can induce flow of the superfluid giving rise to long-lived oscillatory motion of the liquid~\cite{robinson1951adiabatic,hallock1973observation} resulting in thermal instabilities. Similarly, the presence of helium $gas$ in the filling line can give rise to Taconis thermoacoustic oscillations, which can also result in parasitic heat loads~\cite{luck1992thermoacoustic}. These effects are typically combatted by using long, thin tubes (capillary tubes) as filling lines. However, this dramatically increases the flow impedance between the experimental cell and room temperature. This can lead to high pressures when warming the cryostat, a situation exacerbated should accidental air leaks lead to blockages in the filling line. Furthermore, the construction of a complicated filling line that must be made leak-tight to superfluid helium can be a significant drain on the time and patience of the experimentalist.

We have developed a simple, and reliable, hermetically sealed helium source that can be installed directly on the coldest stage of a cryostat, thereby eliminating the need for a filling line. At room temperature, with the cryostat opened, the source cell is charged with helium gas to a desired pressure before being sealed. The cryostat is then closed and cooled as normal, allowing the helium to condense in the source cell and any connected experimental volumes. We demonstrate that this method can successfully supply superfluid helium to a connected experimental cell, using a superconducting microwave resonator device as a sensitive helium detector~\cite{koolstra2019coupling,beysengulov2022helium}. We show that this method eliminates oscillations of the cryostat base temperature arising from the thermomechanical motion of the superfluid, which appear when using a traditional capillary filling line.

Schematic diagrams of the source cell are shown in Fig.~1(a). The source cell is made of OFHC copper and is comprised of a central cylindrical section with upper and lower cap pieces sealed to the cylinder via indium wire o-rings. The internal diameter and length of the cell is 50~mm and 125 mm, respectively; the cell volume is therefore $2.45\times10^5$~mm$^3$ (245~cc). Small demountable brass tube connectors are sealed with indium at the centre of the cap pieces. These brass connectors are soldered to either copper or stainless steel tubing, allowing connections to be made from both the top and bottom. In our experiments, the bottom of the source cell is connected to an experimental cell located beneath base plate of the cryostat by a 200~mm length of thin (1.6~mm outer diameter) stainless steel tubing. 

To prepare the source cell with helium, the top of the cell is connected to a pumping station and a helium gas cylinder via a 3.2~mm diameter copper tube, when the cryostat is open at room temperature. The source cell is alternately evacuated and pressurized with helium several times, to remove air. The pressure of the helium gas in the source cell is then brought to a final value, typically several bar. A hydraulic crimp tool (Solid Sealing Technology KT35046) is then used to make a cold weld crimp seal in the copper tube several cm above the top of the cell. This method has been used previously to make gas gap heat switches and ultra-high quality factor microwave cavities\cite{oriani2022multimodal}. We have found these cold weld seals to be reliably leak tight at room temperature (when testing with a mass spectrometer leak detector) and, more importantly, at low temperatures (by monitoring the cryostat vacuum space pressure whilst flooding the seal internally with superfluid helium). The crimp seal remained leak tight over multiple thermal cycles.

Having thus charged the source cell with helium gas, the cryostat can then be closed and cooled down. The initial pressure of the helium gas in the source cell $P_i$ determines the volume of liquid produced when the cryostat reaches its base temperature. Treating the helium as an ideal gas, the pressure in the source cell decreases linearly with decreasing temperature. Condensation occurs at the temperature for which this pressure becomes equal to the saturated vapor pressure of $^4$He. At this point a fraction of the gas will condense and equilibrium will be established between the liquid and gas phases; further cooling then decreases the vapor pressure, consequently increasing the liquid volume. For $P_i = 4$~bar (absolute) we estimate that condensation occurs at $T = 1.92$~K. Close to 1~K the vapor pressure is negligibly small and we assume that all the helium is in the liquid phase; the final liquid volume is then $\sim$1300~mm$^3$ (1.3~cc).

Before examining source cell performance, we demonstrate that parasitic thermal instabilities can arise when using a filling line to introduce helium into the experimental cell. For these control measurements, a stainless steel filling line, of 1.6~mm outer diameter, runs from room temperature to the experimental cell. This filling line is thermally anchored at each thermal stage of the cryostat by brazing a $\sim$30~cm length of the tube to a copper bobbin that is then clamped onto the thermal plate. The length of tube between each cryostat stage is approximately 30~cm. The cryostat used in our experiments (ICEOxford DRYICE 1.0K) has three thermal stages; the 50K stage and 4K stage, which are cooled by a two-stage Gifford-McMahon cryocooler, and the 1K plate, which is cooled to base temperature ($\sim1.1$~K) by pumping on a helium-4 pot that is continuously filled with helium through a variable needle valve impedance. Once the 1K stage of the cryostat reached base temperature, the experimental cell was filled with 1.0~cc of liquid helium, as determined by the change in pressure in a calibrated external volume at room temperature. In this process, the temperature of each cryostat stage rises slightly as gas cools and condenses into the experimental cell; once all the helium has condensed the temperatures return to their normal values. However, as shown in Fig.~2(a), after several minutes the temperature of the 1K plate begins to oscillate with an amplitude of $\sim10$~mK and a typical period of several minutes.
\begin{figure}[t]
\includegraphics[scale=0.6, angle=0]{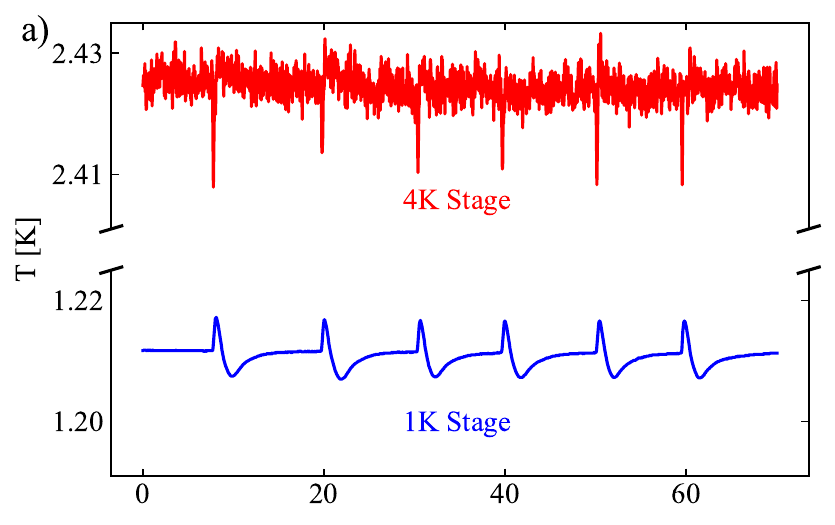}
\includegraphics[scale=0.6, angle=0]{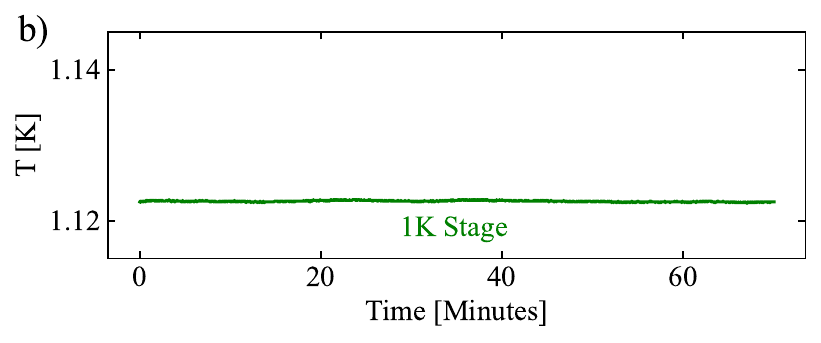}
\caption{\label{fig:epsart} a) Thermoacoustically driven temperature oscillations observed when using a filling line to introduce helium into the experimental cell. b) In contrast, a stable base temperature is observed when using the hermetically sealed source cell to provide helium to the experimental volume. }
\end{figure}
Each maximum in the temperature of the 1~K plate is accompanied by a drop in the temperature of the 4~K plate, indicating a transient increase in the thermal conductance between the stages. Such temperature oscillations are a typical consequence of introducing liquid helium to a multi-stage cryostat; because a van der Waals film (or column of liquid trapped by capillary forces in a thin tube) can propagate upwards in the filling line, the thermal conductance between the stages is increased. Because superfluid helium flows towards the warmer cryostat stages\cite{tisza1938transport} and has extremely high thermal conductivity and vanishingly small viscosity, a \emph{mobile} heat link can be established between the different stages\cite{van1972forces, childers1974observation}. As the superfluid motion is very weakly damped, oscillatory helium flow due to the combined action of thermal and gravitational potential gradients can be established\cite{robinson1951adiabatic}. This motion, combined with the complex interplay between cooling power and transient heat load manifested at each of the cryostat stages, give rise to thermal oscillations detrimental to the experiment under investigation. 

In contrast, when using the hermetic source cell to supply helium to the experimental volume the thermal oscillations are eliminated. In Fig.~2(b) we show a time trace of the cryostat base temperature for the case in which $\sim1.0$~cc of helium is condensed in the source cell and the connected experimental volume. The base temperature remains stable within a window of $\sim1$~mK. This stable performance, and the appearance of temperature oscillations when using the filling line, was observed over multiple cryostat runs and for a variety of experimental conditions~\footnote{We note that during these measurements the filling line used for the control experiments described above remained mounted on the cryostat but was disconnected from the experimental cell and evacuated. The reduced base temperature shown in Fig. 2b) was due to a lower helium flow rate through the 1~K pot. The stable base temperature without helium in the filling line, and the temperature oscillations when helium was present, were observed over a range of base temperatures. For our cryostat the base temperature typically settles within a range of 1.10 to 1.25 K, depending on the experimental conditions.}.  

We now demonstrate the supply of helium from the source cell to the experimental volume of interest. For this purpose, a superconducting resonator device specially designed to be sensitive to thin helium films, as used in several recent studies~\cite{yang2016coupling,koolstra2019coupling, beysengulov2022helium}, was mounted in the experimental cell. 
The device was a `hanger' type resonator patterned on a 100-nm thick niobium film on a silicon wafer as shown in Fig.~3(a). The resonator was capacitively coupled to a coplanar waveguide (CPW) electrode close to one end, and connected to electrical ground at the other. The length and width of the resonator electrode were 5.2~mm and $8.0$~$\mu$m, respectively. The gap between the central resonator electrode and the ground plane was 4~$\mu$m. The exposed silicon in this gap was etched to a depth of 1~$\mu$m to form microchannels running along the length of the resonator electrode. The sample was wire-bonded onto a PCB sample holder that was then mounted inside the experimental cell. Copper spacers were mounted both above and below the PCB to reduce ``box-mode'' resonances within the cell volume. The open volume of the experimental cell was therefore negligible compared with that of the source cell. Hermetically sealed SMP connectors made connection to the PCB allowing measurement of the S$_{21}$ transmission spectrum of the coplanar waveguide with a vector network analyzer. At 1.1~K, and in the absence of helium, a resonant absorption of the microwave signal corresponding to the quarter-wavelength fundamental mode of the resonator was observed at a frequency $f_0 = 6.08$~GHz with loaded quality factor $Q_L = 2.16\times10^4$.
\begin{figure}
\includegraphics[scale=0.1, angle=0]{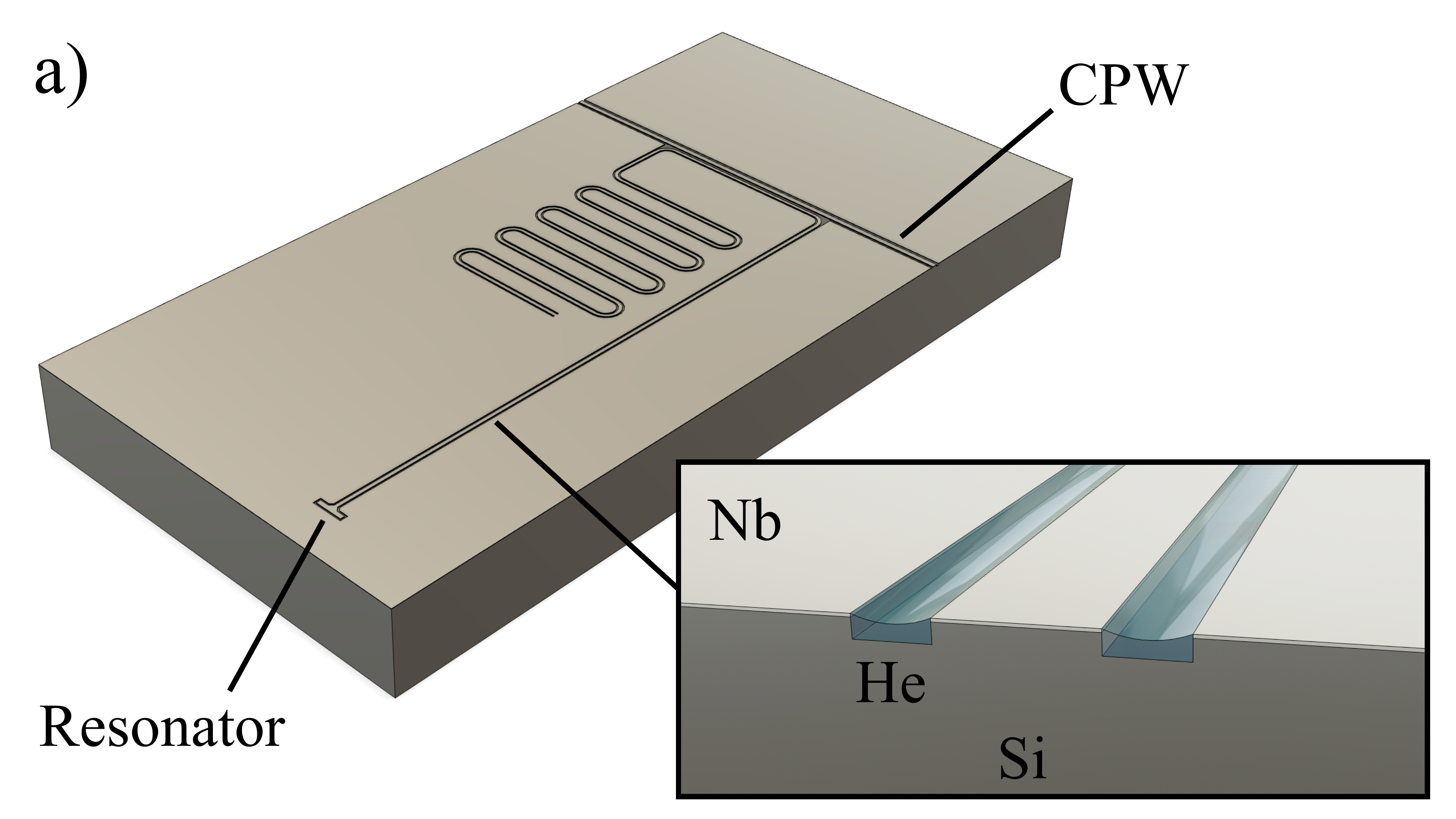}
\includegraphics[scale=0.5, angle=0]{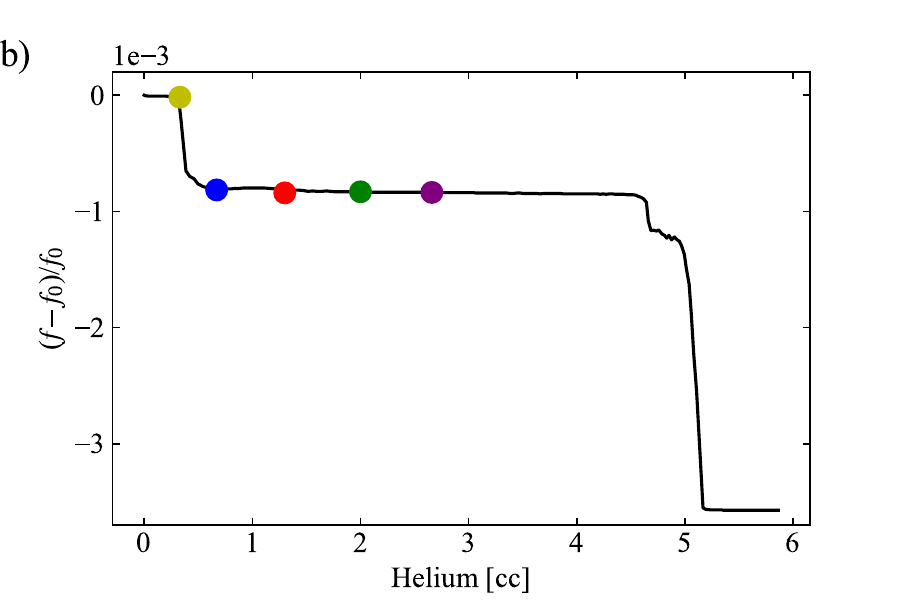}
\includegraphics[scale=0.5, angle=0]{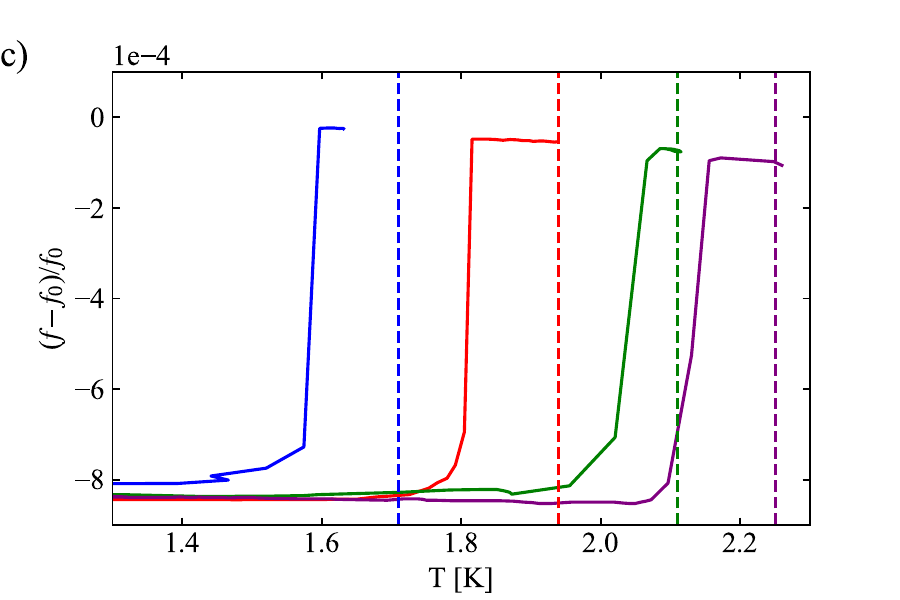}
\caption{\label{fig:epsart} a) Schematic of the superconducting `hanger' resonator device. Inset: Cross-section of the device showing the microchannels formed at the edges of the resonator electrode. b) Solid line: Resonator response upon filling the experimental cell with helium via the filling line. The dots show the frequency shifts observed when employing the source cell with initial loading pressures $P_i = 1, 2, 4, 6$ and 8 bar (yellow, blue, red, green and purple, respectively). c) Resonator response to increasing temperature for $P_i = 2$, 4, 6 and 8 bar. The dashed lines represent the temperature at which the bulk liquid volume in the cell is expected to become fully depleted as the vapor pressure within the experimental cell increases.}
\end{figure}

To calibrate the response of the resonator device to liquid helium, we first used the filling line to slowly introduce helium to the cell while monitoring the resonator frequency $f$. Generally, as liquid helium (dielectric constant $\varepsilon_{\mathrm{He}} = 1.055$) covers the device, $f$ decreases as the change in the dielectric constant local to the resonator electrode reduces the effective speed of light in the device. As shown in Fig.~3(b), and as found in previous studies, this change in frequency occurs in several stages~\cite{koolstra2019coupling, beysengulov2022helium}. First, a van der Waals film is formed on all the surfaces within the experimental cell, including that of the resonator and the adjacent microchannel microstructure. As the film is extremely thin it causes only a very small decrease in the resonator frequency. Then, as bulk liquid begins to collect in the small ($\sim0.2$~cc) open volume at the bottom of the cell beneath the PCB, the channel structures are filled with helium by capillary action of the superfluid~\cite{marty1986stability}. This occurs when the total volume of liquid condensed into the cryostat is approximately 0.5~cc and results in a clear step-like decrease in the resonator frequency of some $8\times10^{-4} f_0$. This shift is in good agreement with finite element modeling (FEM) analysis, which predicts a shift of $-1.0 \times 10^{-3} f_0$ when the channels are completely filled with helium. 

As more helium is introduced into the cell $f$ decreases gradually as the liquid radius of curvature (as determined by the height of the microchannel above the bulk liquid) increases. Eventually however, the amount of helium is sufficient to completely submerge the device resulting in a large frequency shift, observed when $\sim 5$~cc of liquid is condensed into the cell. Again, this frequency shift is in good agreement with FEM analysis, which predicts a shift of approximately $-4.0 \times 10^{-3} f_0$ once the entire volume above the resonator is filled with liquid helium.

This calibration allows us to demonstrate the effectiveness of the source cell in supplying helium to the experiment. In Fig.~3(b) we show values of $f$ recorded at $T=1.1$~K for $P_i=1, 2, 4, 6$ and $8$ bar as the colored dots. The frequency shift values are plotted against the expected volume of liquid formed when all the gas in the source cell condenses. For $P_i=1$ bar, $f$ is very close to the frequency recorded when a thin van der Waals helium film covers the sample, indicating that the amount of helium is not sufficient to form a bulk liquid volume at the bottom of the experimental cell. For $P_i=2$~bar and above, the values are close to the frequency recorded when helium fills the resonator microchannels. This result confirms that the bulk liquid condensed from the source cell collects below the sample allowing the microchannels to fill with superfluid. $P_i \ge $ 15~bar would be required to completely submerge the sample; however, it was found that $P_{i} \ge 10$~bar caused helium gas to leak from the indium seal at the base of the experimental cell. Larger or more robust source cells, or several source cells connected together, could be employed to supply greater volumes of liquid. 
 
Finally, in Fig.~3(c) we show the change in $f$ with temperature $T$ for several values of $P_i$. When $T$ is sufficiently high, the amount of helium is insufficient to maintain the saturated vapor pressure and the bulk liquid is depleted completely. Consequently, the microchannels empty and $f$ jumps suddenly to a value close to $f_0$. For all $P_i$, this frequency jump occurs at a temperature close to, but slightly below, that predicted by the helium vapor-pressure curve. The small discrepancy may reflect that not all the liquid helium in the system collects in the bulk volume, with a fraction ``lost'' to an extended van der Waals film or trapped by capillary action in other small volumes within the cell. 

In conclusion, we have demonstrated a closed-volume helium source that can be charged and sealed \emph{in-situ} on the cold plate of a cryostat. The cold-weld joint used to seal the volume is reliable, robust to superfluid leaks, and simple to make. At low temperature the helium within the cell condenses and becomes superfluid, and can be used for experiments in any connected experimental volume housing quantum devices. This hermetic source cell negates the use of cumbersome external gas handling systems and complicated filling lines mounted within the cryostat, and reduces the heat load to the experiment as well as temperature oscillations arising from the thermomechanical motion of superfluid helium.

We thank A. Oriani for supplying the part number for the hydraulic crimping tool used in this work.

\end{document}